\documentstyle[12pt,psfig]{article}

\newcommand{\AmS}{{\protect\the\textfont2

  A\kern-.1667em\lower.5ex\hbox{M}\kern-.125emS}}

\begin{document}

\centerline{\bf LiBeB: High and low energy cosmic ray production and}
\centerline{\bf comparison with $\nu$ induced nucleosynthesis in SNII}
\vskip 1 truecm

\centerline{Reuven Ramaty}
\centerline{Goddard Space Flight Center, Greenbelt, MD 20771, USA} 
\vskip 0.2truecm

\centerline{Hubert Reeves}
\centerline{Institut d'Astrophysique de Paris, Paris F-75014, France}
\vskip 0.2truecm

\centerline{Richard E. Lingenfelter}
\centerline{University of California San Diego, LaJolla, CA 92093, USA}
\vskip 0.2truecm

\centerline{and}
\vskip 0.2truecm

\centerline{Benzion Kozlovsky}
\centerline{Sackler Faculty of Exact Sciences, Tel Aviv University, Israel}
\vskip 0.2truecm

\begin{abstract}

We present new calculations of LiBeB production by accelerated 
particles with various compositions and energy spectra ranging 
from low energies to relativistic energies, and various ambient 
medium metallicities ($Z/Z_\odot$). The observed, essentially 
constant, Be/Fe ratio as a function of $Z/Z_\odot$ can be best 
understood if the metallicity of the accelerated particles 
(high energy or low energy) does not vary much with 
$Z/Z_\odot$. This could be achieved if the particles are 
accelerated directly from the ejecta of Type II supernovae 
(SNII) and not from the interstellar medium. Using the observed 
Be/Fe and the fact that most of the Fe at low $Z/Z_\odot$ is 
from SNII, we derive the energy content in accelerated 
particles per SNII (a few times 10$^{50}$ erg). We show that 
additional $^{11}$B production by neutrinos is consistent with 
the available data, allowing a neutrino yield from SNII less 
than or equal to the nominal published yields. We further show 
that the observed $^6$Li at low $Z/Z_\odot$ suggests that the 
accelerated particles responsible for the LiBeB at such
$Z/Z_\odot$ are confined to low energies and have a relatively 
high He/O abundance.

\end{abstract}

\vskip 0.5truecm

\noindent{In press:} 
\noindent{Nuclei in the Cosmos 1996, Nuclear Physics, Conference Proceedings}

\eject

\section{INTRODUCTION}

Observations of stars of various metallicities 
($Z/Z_\odot$=(Fe/H)/(Fe/H)$_\odot$) have shown that the 
abundance ratios B/Fe and Be/Fe \cite{ft1} do not vary 
appreciably over a broad range of $Z/Z_\odot$ (see 
\cite{duncan96}). This near constancy implies that the ratios 
of the B and Be production rates to the Fe production rate must 
remain essentially constant as a function of metallicity. 
Moreover, as long as this constancy is maintained and 
destruction effects (due to astration) do not play a major 
role, the production and inventory ratios are equal. For 
$Z/Z_\odot$ less than about 0.1, most of the Fe is thought to 
be produced in Type II supernovae (SNII), with Type I 
supernovae not yet contributing significantly \cite{tt94}. 
Limiting our considerations to this range of metallicities, the 
constancy of B/Fe and Be/Fe strongly suggest that B and Be are 
also produced by SNIIs. The accepted paradigm for the origin of 
most of the Be is production in spallation reactions of 
accelerated particles ($p$, $\alpha$, CNO) deriving their 
energy from supernovae. B, in addition to such reactions, is 
also produced in neutrino interactions in SNIIs 
\cite{wh90,ww95,vf96}. The observed Be abundance as a function 
of metallicity can be approximated by 
Be/H$\simeq$5.2$\times$10$^{-11}$(Fe/H)/(Fe/H)$_\odot$ 
\cite{duncan96}, implying that 
Be/Fe$\simeq$1.7$\times$10$^{-6}$. The relatively well known Fe 
yield per SNII ($\sim$0.1M$_\odot$ based on SN87A observations) 
then provides a direct determination of the accelerated 
particle-produced Be per SNII 
($\sim$3.6$\times$10$^{48}$atoms), independent of the details 
of the spallation reactions. The total energy per SNII imparted 
to the accelerated particles to produce the required amount of 
Be, as well as the accompanying production of the other light 
isotopes, do depend on the accelerated particle composition and 
spectrum, the ambient medium composition and the path length 
$X_{esc}$ available to the accelerated particles for producing 
nuclear reactions.

Previous calculation of LiBeB production by accelerated 
particles \cite{fields94,rkl96} incorporated various 
compositions and energy spectra but used different codes for 
spectra extending to relativistic energies (hereafter HECR) or 
spectra confined to low energies (hereafter LECR). Furthermore, 
the HECR calculations at epochs of low metallicity implicitly 
assumed that the composition of the HECR is the same as that of 
the ambient medium \cite{st92,pr93,fields94}. In addition, secondary 
reaction chains (e.g. fast $^{12}$C on ambient H producing 
$^{11}$B and the subsequent interaction of the fast $^{11}$B to 
produce $^{10}$B or lighter isotopes), which are quite 
important when HECR with solar or higher metallicity are 
interacting with a low metallicity ambient medium, have been 
ignored. We have developed a new code that employs updated 
cross sections \cite{rv84,w90,w96} and incorporates all the 
possible secondary chains for source particles up to $^{16}$O. 
We calculate LiBeB production for a variety of accelerated 
particle spectra (LECR to HECR), ambient medium metallicities 
10$^{-3}$$<$$Z/Z_\odot$$<$1, and various         accelerated 
particle compositions. By normalizing the LiBeB yields to the 
energy content of the source accelerated particles, we can 
directly determine the SNII particle acceleration efficiency 
required to produce the observed Be/Fe. We next combine the 
accelerated particle-produced and $\nu$-produced $^{11}$B and 
set constraints on the contribution of the $\nu$ process on B 
production. Finally, we consider the implications of the $^6$Li 
observations at low metallicity \cite{sln93}.

\section{ANALYSIS}

We assume that the ambient composition is solar scaled by a 
variable metallicity. For the fast particle composition we use 
the following: (i) ejecta of Type II supernovae (12-40 
M$_\odot$ progenitors) for various metallicities \cite{ww95} 
averaged over an IMF proportional to $M^{-2.7}$ (hereafter 
SNII); (ii) current epoch cosmic ray source abundances 
\cite{en90,lu89} (hereafter CRS); (iii) ambient medium 
abundances, i.e. solar system abundances scaled with 
metallicity (hereafter SSz); and (iv) the ejecta of a supernova 
from a 60 M$_\odot$ progenitor \cite{wlw93} (hereafter SN60). 
The SNII abundances of the most relevant isotopes, $^1$H, 
$^4$He, $^{12}$C and $^{16}$O, are weakly dependent on 
$Z/Z_\odot$ and are approximately proportional to 100:20:0.2:1, 
respectively. The CRS and SN60 abundances are taken independent 
of $Z/Z_\odot$ and the above proportions are 175:25:0.8:1 and 
0:0.7:0.5:1, respectively. Thus, CRS is not too different from 
SNII except that it is more abundant in $^{12}$C; for SN60, H 
and He are absent or very under abundant, and, as for CRS, 
$^{12}$C is more abundant than for SNII. We employ a shock 
acceleration spectrum with a high energy cutoff, 
$(p^{-s}/\beta) {\rm e}^{-E/E_0}$, except for SN60 for which, 
as before \cite{rkl96}, we assume that above 50 MeV/nucl the 
spectrum decreases as $E^{-10}$; $p$ and $E$ are momentum and 
kinetic energy, both per nucleon, and $c\beta$ is particle 
velocity. The SN60 model, because of its relatively high C/O 
and sharp high energy cutoff, produces the highest B/Be and 
$^{11}$B/$^{10}$B without violating energy constraints (see 
below). We assume that $X_{esc}$ is energy independent, and 
present results for $X_{esc}$=10 gcm$^{-2}$, a leaky box model, 
and $X_{esc} \rightarrow \infty$, a closed galaxy model. The 
neglect of the energy dependence affects our results only 
marginally. The incorporation of the secondary reaction chains 
can lead to contributions as high as 60\% of the total 
production in some cases. 

The nuclear reactions are accompanied by energy loss due to 
Coulomb interactions of the accelerated particles with the 
ambient medium (assumed to be neutral). Fig.~1 shows the Be 
production per deposited energy. The solid curves are for SNII, 
the dashed line is for SSz and the dashed line with diamonds is 
for SN60. The values for CRS are practically identical to those 
for SNII with $E_0$=10GeV/n. 
All curves with given values of $E_0$ are for the 
exponential cutoff at the indicated $E_0$. We see that for  
SNII and SN60, for which the metallicities of the accelerated 
particles are nearly independent of $Z/Z_\odot$, $Q(Be)/W$ is 
practically independent of the ambient medium metallicity; 
this, however, is not the case for SSz for which the 
metallicity of the fast particles increases with increasing 
$Z/Z_\odot$. $Q(Be)/W$ decreases rapidly for $E_0<50$MeV/n; we 
show values as a function of $Z/Z_\odot$ for $E_0$=10MeV/n. 
Using the values plotted in Fig.~1 we can express the Be yield 
per SNII as $Q(Be)=10^{51}W_{51}[Q(Be)/W)]$, where 
10$^{51}W_{51}$ erg is the energy content in accelerated 
particles per SNII. As the total SNII mechanical energy is 
$\sim$10$^{51}$ erg, $W_{51}$ is essentially the acceleration 
efficiency. We assume that $W_{51}$ is independent of 
$Z/Z_\odot$.

The calculated \cite{ww95} $^{56}$Fe yields, averaged over the 
IMF given above, range from about 0.09 to 0.14 M$_\odot$ per 
SNII. As already indicated, we adopt the constant value of 0.1, 
leading to a required Be production per SNII of 
3.6$\times$10$^{48}$ atoms independent of $Z/Z_\odot$. Then for 
the SNII, CRS and SN60 compositions, for which $Q(Be)/W$ is 
independent of $Z/Z_\odot$, 
$W_{51}[Q(Be/W)]$$\simeq$3.6$\times$10$^{-3}$. Requiring an 
acceleration efficiency less than about 0.3, 
$Q(Be)/W$$\ge$1.2$\times$10$^{-2}$. This effectively rules out 
cutoff energies $E_0$$<$50 MeV/n, as well as the SSz case. The 
latter is also ruled because of its dependence on $Z/Z_\odot$.

Fig.~2 shows $Q(B)/Q(Be)$, the accelerated particle produced 
B-Be ratio, for all cases except SSz. $Q(B)/Q(Be)$ 
is practically independent of $Z/Z_\odot$, 
implying that (B/Be)$_{cr}$ can remain constant as a function 
of $Z/Z_\odot$. The calculated $Q(B)/Q(Be)$ ranges from about 
10 to 30, depending on the accelerated particle spectrum and 
composition. $^{11}$B can also be produced by $\nu$ spallation 
in supernovae \cite{wh90,ww95}. By averaging the calculated 
$^{11}$B yields over the same IMF, we obtain a $\nu$-induced 
$^{11}$B production of $\simeq$6$\times$10$^{-7}$M$_\odot$ per 
SNII, practically independent of metallicity. For the total 
B-to-Be inventory ratio we thus have 

\begin{equation}
{\rm B/Be} \simeq ({\rm B/Be})_{cr}+3.0{\rm M}_{7}, 
\end{equation}

\noindent where M$_{7}$ is the $\nu$-produced $^{11}$B in units 
of 10$^{-7}$M$_\odot$. The second term in eq.~(1) is simply 
determined by the numbers of $^{56}$Fe and  $\nu$-produced 
$^{11}$B per SNII, and the observed Be/Fe ratio. The observed 
B/Be is quite uncertain, mainly because of NLTE effects. 
Adopting the range 9$<$B/Be$<$34 \cite{edv94} for low 
metallicity stars, the maximum contribution of $\nu$ 
spallation, M$_{7}$$\simeq$8, is obtained for the minimal 
(B/Be)$_{cr}$$\simeq$10 corresponding to production by 
accelerated particles extending to GeV energies in the closed 
galaxy model. This upper limit slightly exceeds the predicted 
$^{11}$B production by neutrinos \cite{ww95}. For the other 
accelerated particle compositions, $Q(B)/Q(Be)$ is larger and 
consequently the maximal M$_{7}$ is smaller. For example, for 
SNII with $E_0=50$MeV/n, $Q(B)/Q(Be)$$\simeq$18 and hence 
M$_{7}$$\le$5, a limit which is slightly lower than the 
predicted value.

Fig.~3 shows $Q(^{11}B)/Q(^{10}B)$ together with the meteoritic 
B isotopic ratio of about 4. This isotopic ratio cannot be 
reproduced by HECR which yield an isotopic ratio less than 
about 2.5 (Fig.~3). The addition of the $\nu$-produced $^{11}$B 
can provide the extra $^{11}$B. Using eq.~(1) we obtain 
$^{11}$B/$^{10}$B= [3M$_7$/(B/Be)$_{cr}$] 
[($^{11}$B/$^{10}$B)$_{cr}$+1] +($^{11}$B/$^{10}$B)$_{cr}$, 
where $^{11}{\rm B}/^{10}{\rm B}$ includes both the accelerated 
particle and $\nu$-produced $^{11}$B. If for the low 
metallicity stars $^{11}{\rm B}/^{10}{\rm B}$ equals the solar 
system value of 4, then for the various models 0$<$M$_7$$<$2, 
i.e. ranging from a negligible $\nu$ contribution to less than 
about 30\% of the predicted value.

Fig.~4 shows $Q(^{6}Li)/Q(Be)$. For the compositions that we 
use, this ratio also remains constant as a function of 
$Z/Z_\odot$ allowing us to compare production and inventory 
abundance ratios. A value of 0.05$\pm$0.02 for $^6$Li/Li was 
observed at $Z/Z_\odot\simeq10^{-2.4}$ \cite{sln93}. The error 
bar (SLN) shown at this metallicity is based on this 
observation, and on the observed Li/H (Spite plateau, error 
$\pm$47\%) and Be/H at $Z/Z_\odot\simeq10^{-2.4}$, for which we 
estimate an error of $\pm$50\%. We see that SNII with 
$E_0$$\simeq$50MeV/n produces more $^6$Li relative to Be than 
both HECR and SN60. HECR underproduce $^6$Li because the 
$\alpha-\alpha$ cross section is relatively low at high 
energies; and SN60 underproduces $^6$Li because of the low 
$\alpha$ particle abundance. Although the large error bar of 
the SLN data does not allow us to discriminate between the 
models, these data favor SNII with $E_0$$\simeq$50MeV/n rather 
than either HECR or SN60. On the other hand, the meteoritic 
$^6$Li/Be at $Z/Z_\odot$=1 (Fig.~4) is consistent with these 
models suggesting that the LiBeB at the time of the formation 
of the solar system resulted from more than one accelerated 
particle component. We mention in addition that the high 
$Q(^{6}Li)/Q(Be)$ implied by SNII with $E_0$$\simeq$50MeV/n 
tends to overproduce Li/H at $Z/Z_\odot$ greater than about 
0.1. To overcome this problem, the SN60 model was used in a 
recent galactic abundance evolution study \cite{vf96}. But as 
we see from Fig.~4, SN60 undeproduces $^6$Li at low 
metallicity. Our conclusion concerning this $^6$Li differs from 
a previous one \cite{sln93}, namely   that the observed $^6$Li 
abundance can be accounted for by HECR spallation. That 
suggestion was based on the assumption that the composition of 
the HECR is the same as that of the interstellar medium at low 
metallicity (i.e. SSz), in which case the high He to CNO ratio 
for the accelerated particles would indeed yield a high 
$^6$Li/Be. But, as we have seen, such a composition fails 
energetically to account for the Be abundance and does not 
yield a constant Be/Fe ratio.

\begin{figure}[htb]

\begin{minipage}[t]{102mm}
\psfig{file=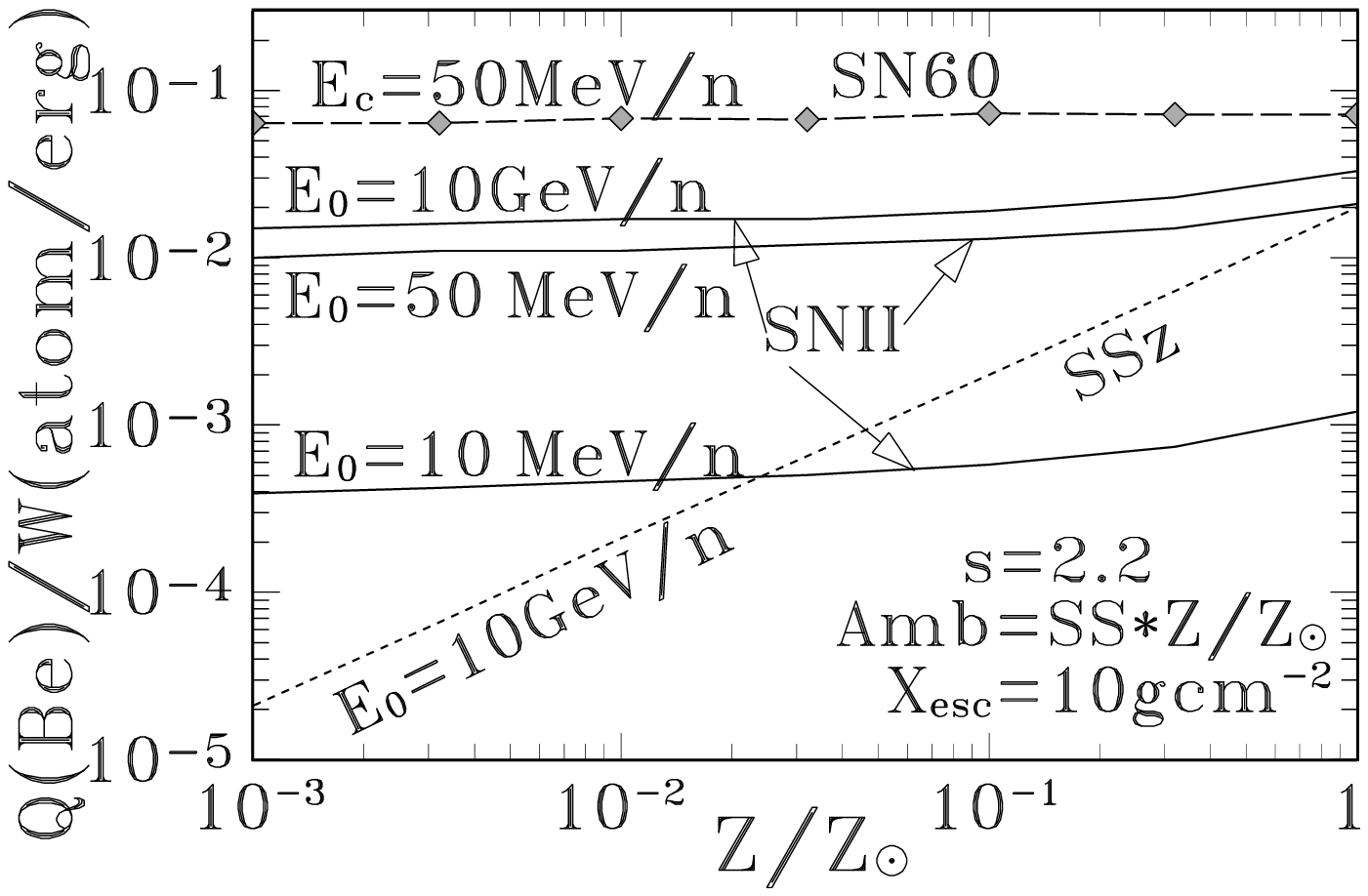,width=100mm}
\vskip -0.3truecm
\caption{Number of Be atoms produced per deposited erg. The 
ambient medium has solar composition scaled with metallicity 
(SSz).  For all curves except SSz the metallicity of the 
accelerated particles does not change much with $Z/Z_\odot$.}
\end{minipage}
\hspace{\fill}
\begin{minipage}[t]{102mm}
\vskip 0.3truecm
\psfig{file=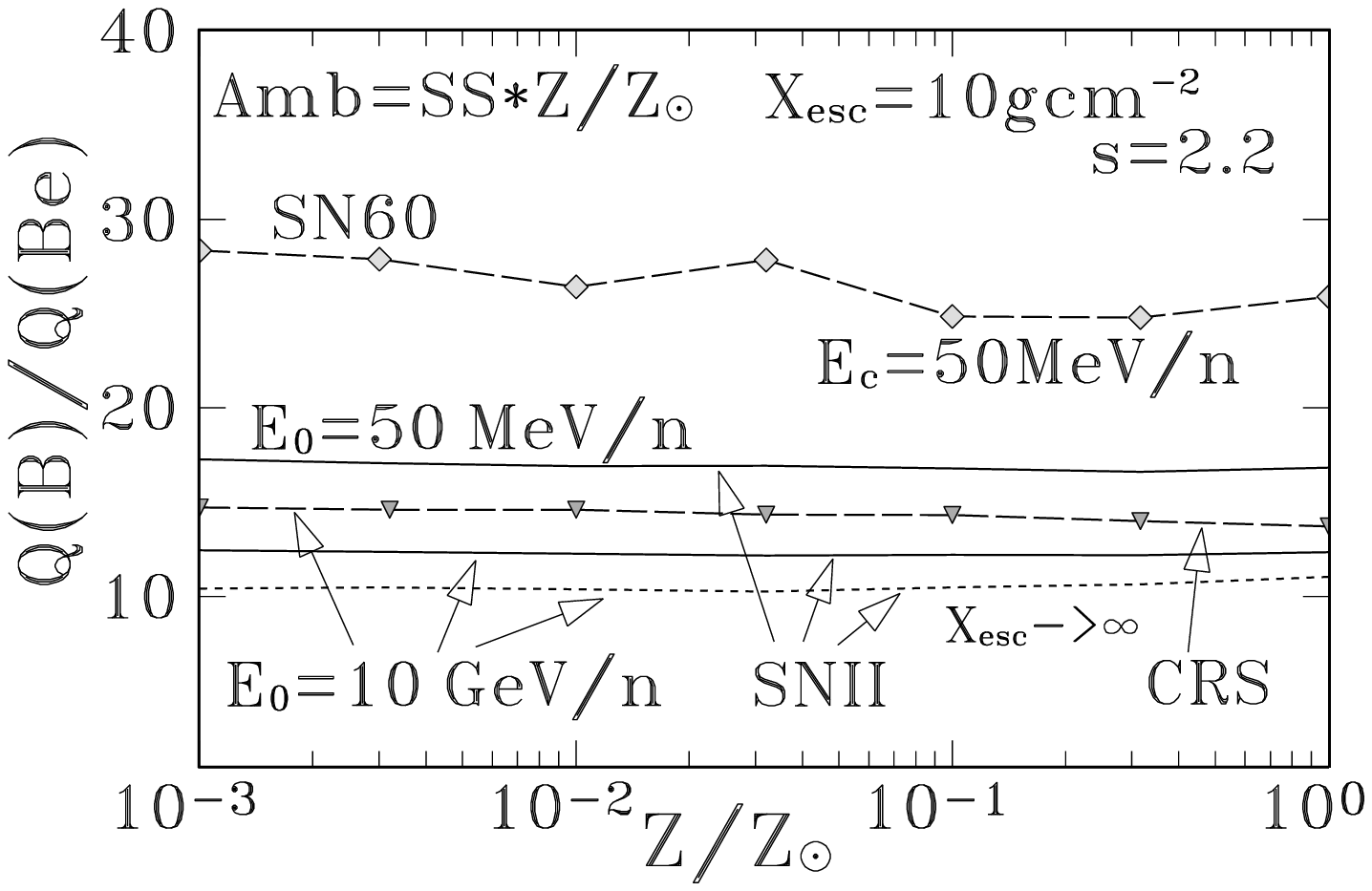,width=100mm}
\vskip -0.3truecm
\caption{B-Be production ratio by accelerated particle interactions. The 
ambient medium and accelerated particle compositions are as in 
Fig.~1, except that we do not show the SSz composition. The 
lowest (dashed) line is for the closed galaxy model.}
\end{minipage}

\end{figure}

\begin{figure}[htb]

\begin{minipage}[t]{102mm}
\psfig{file=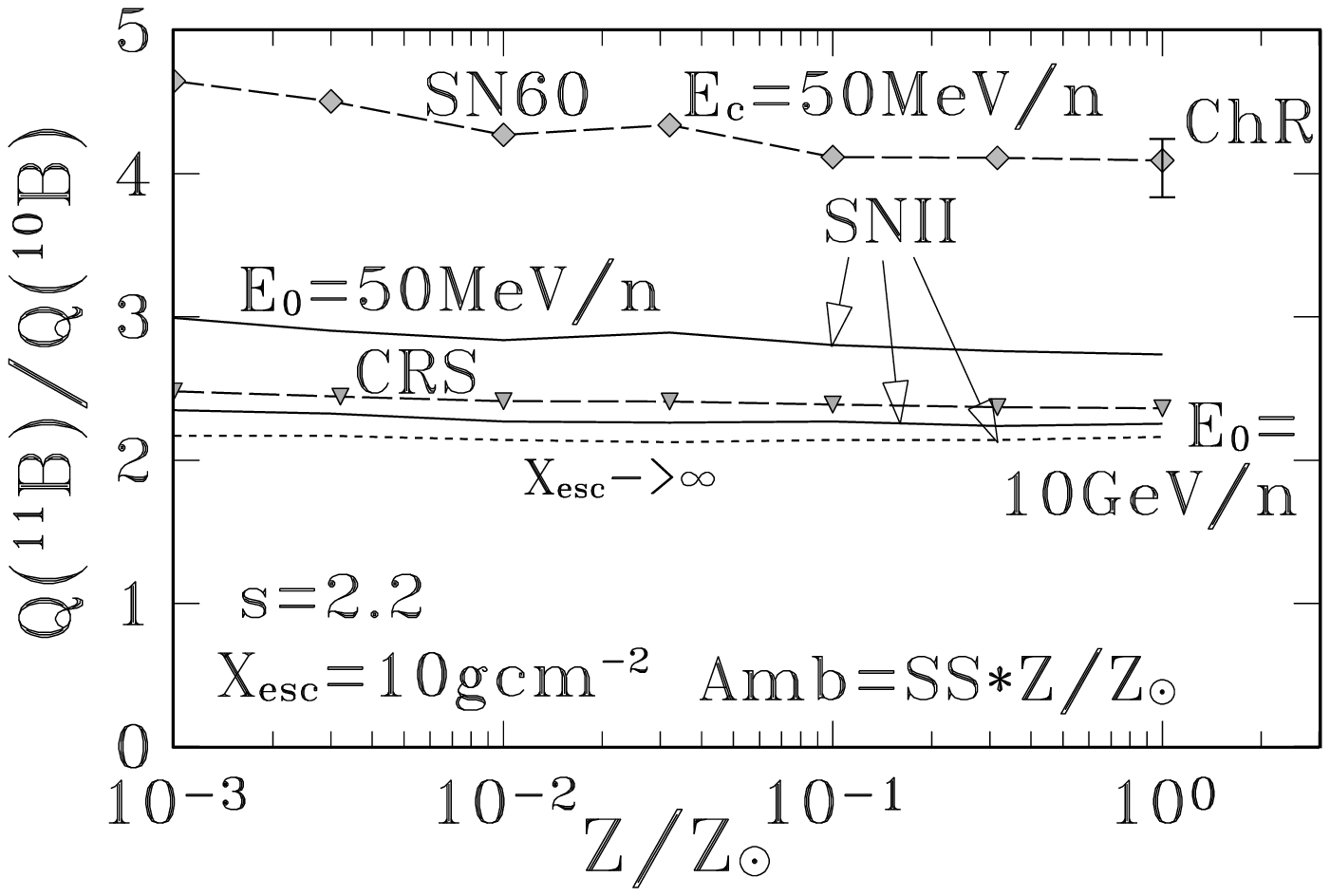,width=100mm}
\vskip -0.3truecm
\caption{B isotope production ratio. SN60 achieves values $>$4 
because of its relatively high C/O and sharp cutoff at 50 MeV/n.}
\end{minipage}
\hspace{\fill}
\begin{minipage}[t]{102mm}
\vskip 0.3truecm
\psfig{file=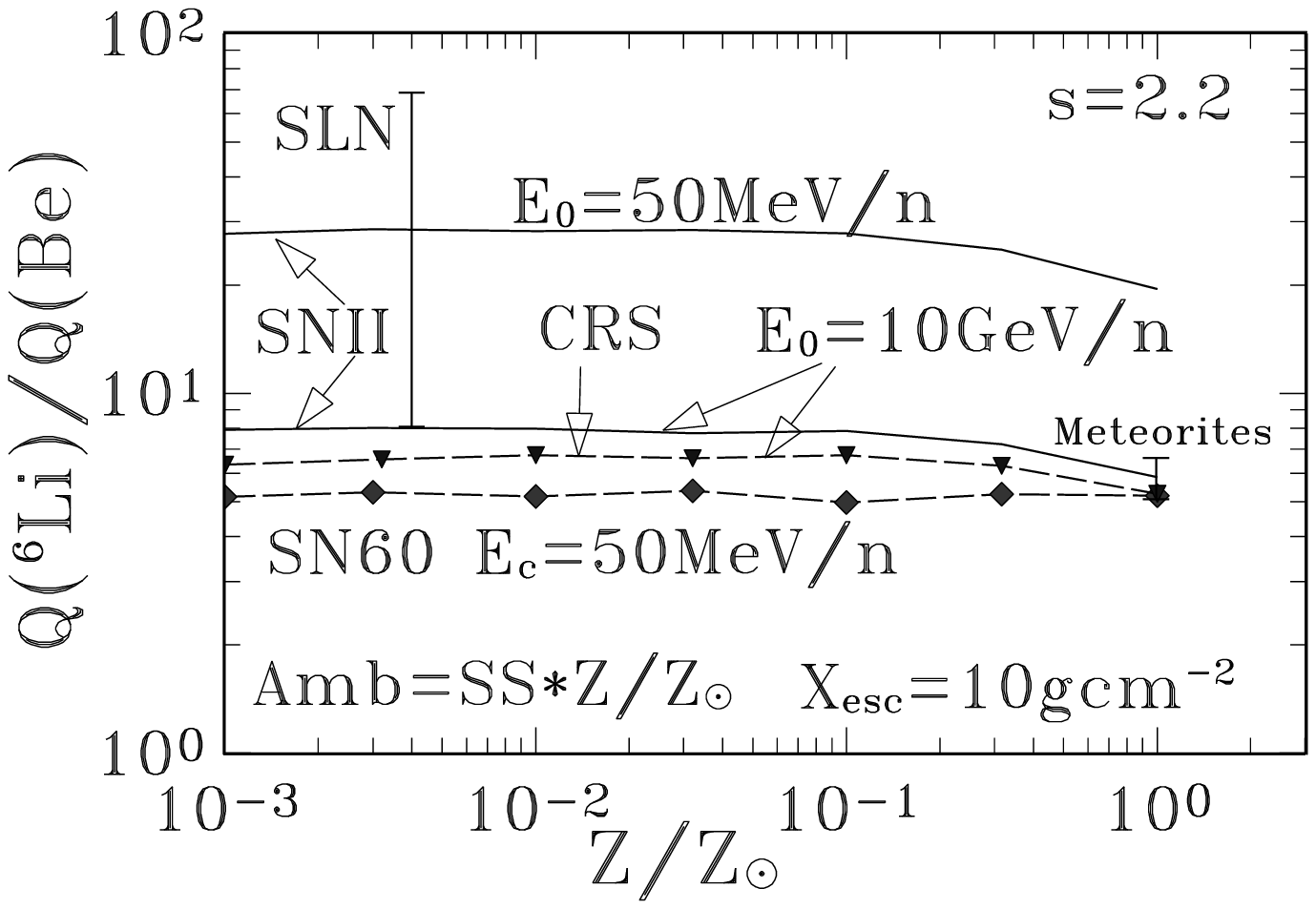,width=100mm}
\vskip -0.3truecm
\caption{$^6$Li-Be production ratio. SLN  represents the 
observed [17] $^6$Li/Li combined with the Li/H and Be/H data.} 
\end{minipage}

\end{figure}


\begin{thebibliography}{9}
\bibitem{ft1} We adopt the convention of using ratios of 
chemical or isotopic symbols to denote ratios by number of the 
galactic inventories of the respective elements or isotopes.
\end{thebibliography}
\end{document}